\documentclass[aps,prb,twocolumn,superscriptaddress,oneside,floatfix,amsmath,showpacs,amssymb]{revtex4}
\usepackage{graphicx}
\usepackage{dcolumn}
\usepackage{bm}
\usepackage[usenames]{color}
\usepackage{doi}
\usepackage{hyperref}

\begin{document}

\newcommand{\etal}{\textit{et al.}}


\title{Magnetoplasmon resonances in polycrystalline bismuth as seen via terahertz spectroscopy}

\author{Julien Levallois}
\email{julien.levallois@unige.ch}
\affiliation{D\'{e}partement de Physique de la Mati\`{e}re Condens\'{e}e, Universit\'{e} de Gen\`{e}ve, CH-1211 Gen\`{e}ve 4, Switzerland}

\author{Piotr Chudzi\'{n}ski\footnote{Present address: Institute of Theoretical Physics, University of Regensburg, D-93040 Regensburg, Germany} }
\email{Piotr.Chudzinski@physik.uni-regensburg.de}
\affiliation{D\'{e}partement de Physique de la Mati\`{e}re Condens\'{e}e, Universit\'{e} de Gen\`{e}ve, CH-1211 Gen\`{e}ve 4, Switzerland}

\author{Jason N. Hancock\footnote{Present address: Department of Physics and Institute of Materials Science, University of Connecticut, Storrs, Connecticut 06269} }
\affiliation{D\'{e}partement de Physique de la Mati\`{e}re Condens\'{e}e, Universit\'{e} de Gen\`{e}ve, CH-1211 Gen\`{e}ve 4, Switzerland}

\author{Alexey B. Kuzmenko}
\affiliation{D\'{e}partement de Physique de la Mati\`{e}re Condens\'{e}e, Universit\'{e} de Gen\`{e}ve, CH-1211 Gen\`{e}ve 4, Switzerland}

\author{Dirk van der Marel}
\affiliation{D\'{e}partement de Physique de la Mati\`{e}re Condens\'{e}e, Universit\'{e} de Gen\`{e}ve, CH-1211 Gen\`{e}ve 4, Switzerland}

\date{\today}


\begin{abstract}

We report the magnetic field-dependent far-infrared reflectivity of polycrystalline bismuth. We observe four distinct absorptions that we attribute to magnetoplasmon resonances, which are collective modes of an electron-hole liquid in magnetic field and become optical and acoustic resonances of the electron-hole system in the small-field limit. The acoustic mode is expected only when the masses of distinct components are very different, which is the case in bismuth. In a polycrystal, where the translational symmetry is broken, a big shift of spectral weight to acoustic plasmon is possible. This enables us to detect an associated plasma edge. Although the polycrystal sample has grains of randomly distributed orientations, our reflectivity results can be explained by invoking only two, clearly distinct, series of resonances. In the limit of zero field, the optical modes of these two series converge onto plasma frequencies measured in monocrystal along the main optical axes.

\end{abstract}


\maketitle


\section{Introduction}
\begin{figure}[t]
\centering
\includegraphics[width=\columnwidth]{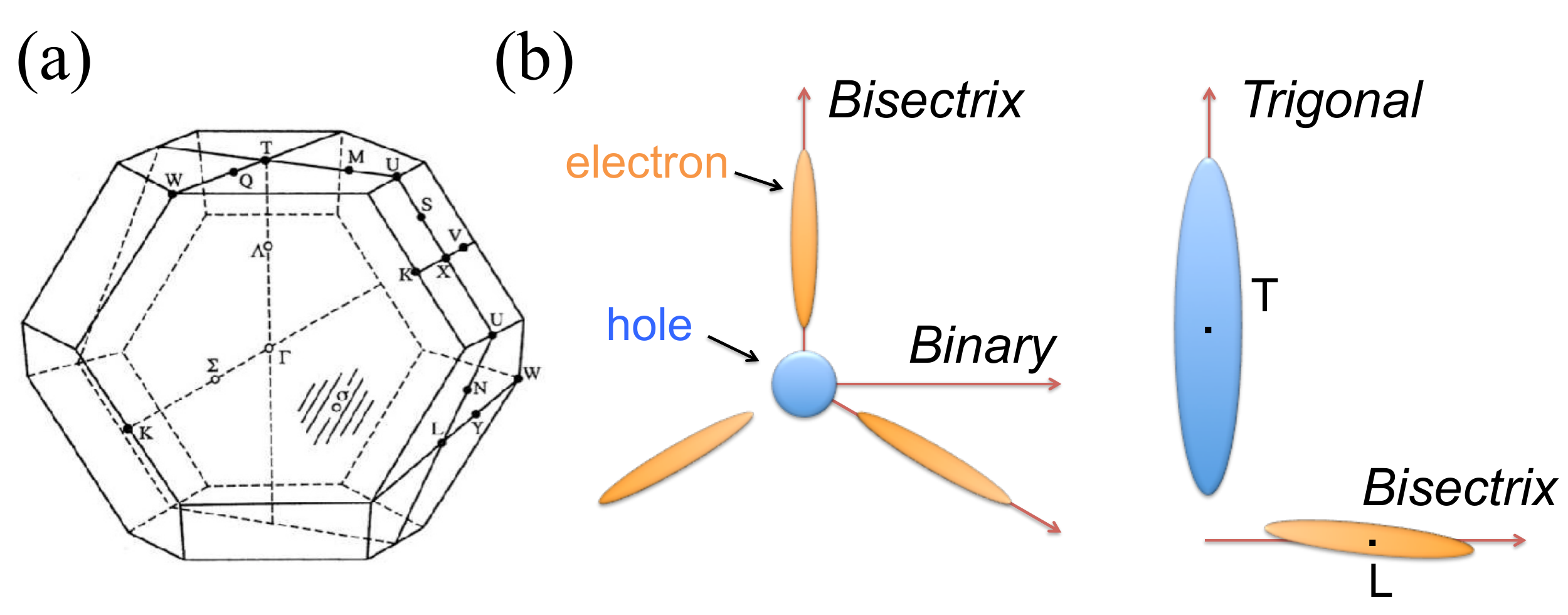}
\caption{(Color online) a) The Brillouin zone with the main
symmetry points (adapted from Ref.~\onlinecite{Liu95}). b) The Fermi
surface of bismuth: a hole pocket which is an ellipsoid elongated
along the trigonal axis (C$_3$) and three electron pockets
distributed in a threefold degenerate pocket around the C$_3$
axis. They are not perfectly aligned with the bisectrix axis
(C$_1$) but tilted by 6$^\circ$. The binary axis (C$_2$) is
orthogonal to the C$_1$ axis.} \label{Bismuth}
\end{figure}

Pure bismuth exhibits rich physics due to its particular electronic properties. In its pristine state, it is an exactly charge compensated semimetal of electrons and holes. Peculiar to bismuth is the low carrier density ($\sim$10$^{17}$cm$^{-3}$) and extremely anisotropic cyclotron masses~\cite{Zhu11}. Magneto-oscillation phenomena are therefore quite easily accessible and in fact were discovered in bismuth~\cite{Haas30,Shubnikov30}. The Fermi surface of bismuth, which is depicted in Fig.~\ref{Bismuth}, is highly anisotropic and made by three elongated ellipsoid (Dirac) electron pockets, which are slightly tilted by 6$^\circ$ from the bisectrix axis (C$_1$), and one ellipsoid hole (massive) pocket along the trigonal axis (C$_3$)~\cite{Liu95}, each having volume which represents only few percent of the Brillouin zone. This material is of central interest because it is a key player of high-T$_c$ superconductors, it has been found to host a transition to a topological insulating state by partial substitution with iso-electronic antimony~\cite{Hsieh08} and is promising for valley-tronics~\cite{Zhu11np}. The observations of a valley-ferromagnetic state~\cite{Li08} and optical evidence for electron-plasmon coupling~\cite{Tediosi07} have lead to a revival of interest in the collective excitations and low-temperature transport properties of this material~\cite{Piotr11}.


The extremely large ratio of masses of different carriers has one highly non-trivial implication: the presence of an additional collective excitation of a Fermi liquid, an acoustic resonance. Its existence in two component plasma has been suggested already long time ago~\cite{Pines62}, but the collective mode turned out to be quite elusive. In particular from the viewpoint of optical spectroscopy there are two fundamental obstacles. Firstly, when $q\rightarrow 0$, which is the usual resonance condition with photons, the energy of acoustic plasmon vanishes. This issue can be solved by applying a magnetic field because the movement of a charge under a magnetic field requires an extra kinetic energy to overcome the magnetic field vector potential; then the resonance frequency becomes $\omega=\sqrt{\omega_{p}^2+\omega_{c}^2}$, where $\omega_p$ and $\omega_c$ are the screened plasma and the cyclotron frequency respectively. The reason why such simple addition formula works is thanks to Kohn's theorem and has been shown to work in numerous studies: from a semi-classical approximation for dielectric tensor~\cite{semi-class-1st-Bi} or a self-consistent field~\cite{Mermin-self-consis}, later with equation of motion technique~\cite{Quinn-RPA-eq-of-motion}, finally using polarizabilities known from random-phase approximation (RPA) polarizability~\cite{Noto74} and recently extended by including vertex corrections to RPA~\cite{Halperin-extend-diagram}.

Secondly, in the case of translationally invariant systems, the longitudinal $f$-sum rule warrants that in the limit of small fields $B\rightarrow 0$ the excitation proportional to the total carrier density (standard, optical plasmon) exhausts all the spectral weight. This creates another serious obstacle in experimental efforts to directly observe acoustic plasmons because these low energy collective excitations would have a noticeable spectral weight only at rather high magnetic field (for Bi above $\approx$ 6~T) when it will slowly turn into cyclotron resonance of the heavier carriers (see Eq.~\ref{plasmon} and discussion in Sec.~\ref{sec:theo-disc}).

The translational invariance, or to be more precise, the fact that electron wavelength is a good quantum number, is a necessary condition used in every derivation of the above mentioned longitudinal $f$-sum rule. Thus, the natural way to overcome this obstacle and detect an acoustic plasmon is to work with a system with a broken translational symmetry. This is our motivation to perform a measurement on a bismuth polycrystal.%
\section{Experimental}\label{sec:method}
The sample used in this study is a bismuth polycrystal which consists of half a disk of 12~mm diameter and 2~mm thick. The sample was characterized by X-ray diffraction which shows that all orientations (C$_1$, C$_2$ and C$_3$) are equally and randomly present with grain size of $\sim$~100~nm.
We have measured the optical reflectivity at $\sim$5~K in the range 5 to 650~cm$^{-1}$ by time-domain THz spectroscopy (TPI spectra 1000, TeraView Ltd.) using a home-made (magneto-)reflectivity setup up to 3.4~T and by Fourier transform infrared spectroscopy up to 7~T. In the latter case, the reflectivity was obtained using a gold mirror reference. For both instruments the magnetic field was applied perpendicular to the surface of the sample (Faraday geometry) and nearly parallel to the propagation of light. The Kerr angle is defined as the rotation of the polarization plane upon a normal reflection of linearly polarized light from the surface of a sample and with the magnetic field B perpendicular to the surface. The Kerr angles were determined for each frequency separately by examining the dependence on analyser angle, as described in detail in Levallois \etal~\cite{Levallois12}.


\section{Results}\label{sec:result}


\begin{figure}[t]
\centering
\includegraphics[width=\columnwidth]{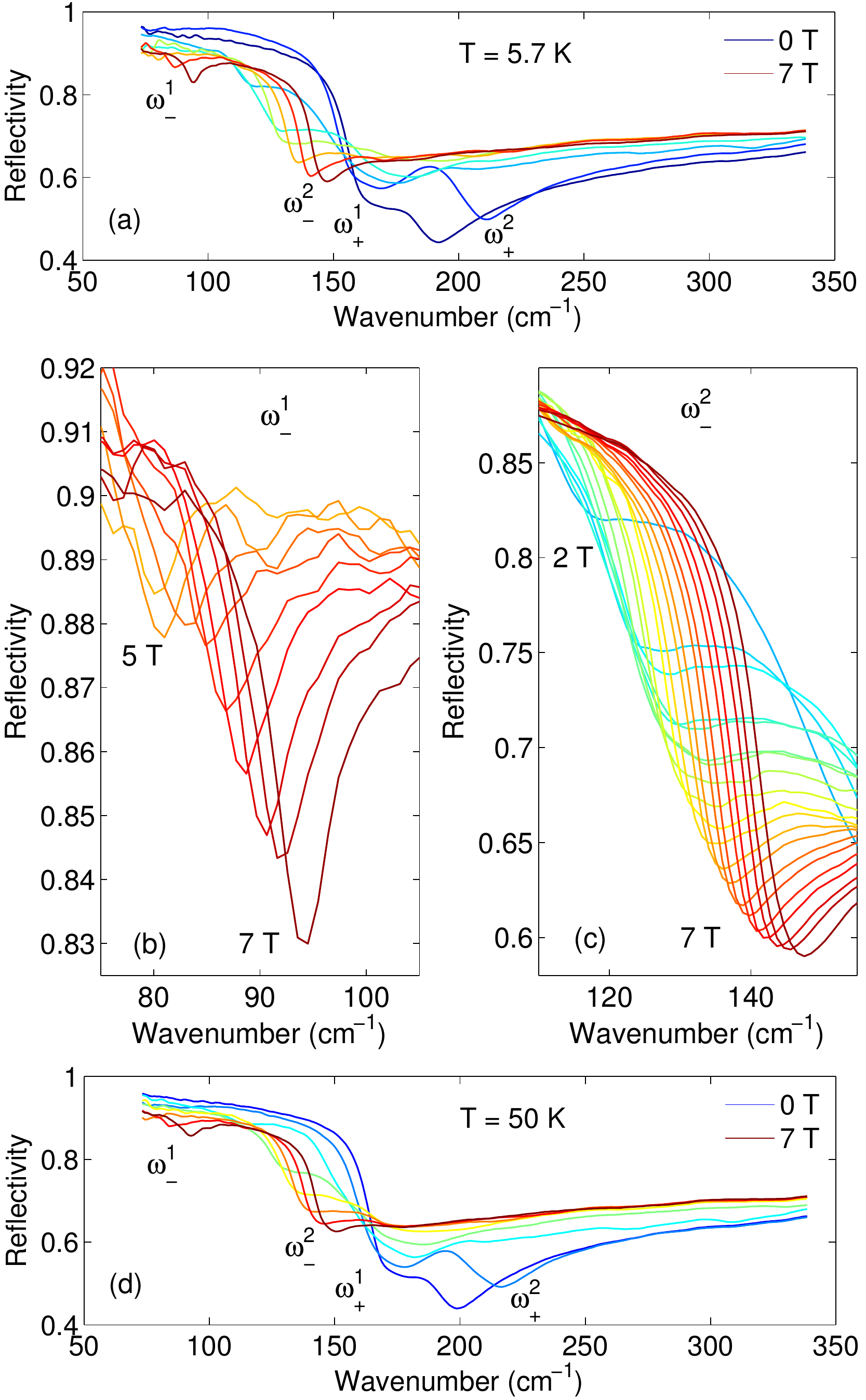}
\caption{(Color online) a) Far infrared reflectivity of polycrystalline bismuth
at 5.7~K as a function of frequency at various magnetic fields (each tesla).
b) and c) Zoom of the reflectivity around 90~cm$^{-1}$ in order to highlight
$\omega_-^1$ and around 130~cm$^{-1}$ in
order to highlight $\omega_-^2$, for every 0.25~T between 0 and 7~T. d) Reflectivity at 50~K as a function of frequency at
various magnetic fields (each tesla).}
\label{data}
\end{figure}

The reflectivity of polycrystalline bismuth at 5.7~K as a function of frequency and field is presented in Fig.~\ref{data}a. The general shape is in good agreement with the one of a single crystal oriented along the trigonal axis~\cite{Tediosi07}. The reflectivity exhibits a very strong field dependence and we clearly observe four different plasma edges in spectra collected at high field. As the magnetic field increases, the two plasma edges which are present at zero field, labelled $\omega_+^1$ and $\omega_+^2$, shift to higher frequencies. Whereas $\omega_+^1$ is present up to 6.5 T, $\omega_+^2$ disappears above 2~T. At lower frequency, we observe the apparition of two other structures, labelled $\omega_-^1$ and $\omega_-^2$, whose positions and intensities increase with magnetic field, as shown in Fig.\ref{data}b,c. The strength of those absorptions, which is proportional to the spectral weight of a given bosonic mode, increases quite linearly at low magnetic field, as suggested in Ref.~\onlinecite{Piotr11}.

The four resonances are also visible at higher temperature, as shown in Fig.~\ref{data}d where the reflectivity spectra at 50~K at various magnetic fields  are presented. We can see that at 50~K the modes $\omega_+^1$ and $\omega_+^2$ disappear around 4~T and 1~T respectively, \emph{i.e.} at lower fields than at 5~T. For these field values these magneto-plasma mode shifts above 220 cm$^{-1}$, which is also the lower bound of the interband optical conductivity seen in single crystal data~\cite{Tediosi07}, suggesting that Landau-damping is the main reason for the disappearance of the optical magnetoplasmons at high magnetic fields.


\begin{figure}[t]
\centering
\includegraphics[width=\columnwidth]{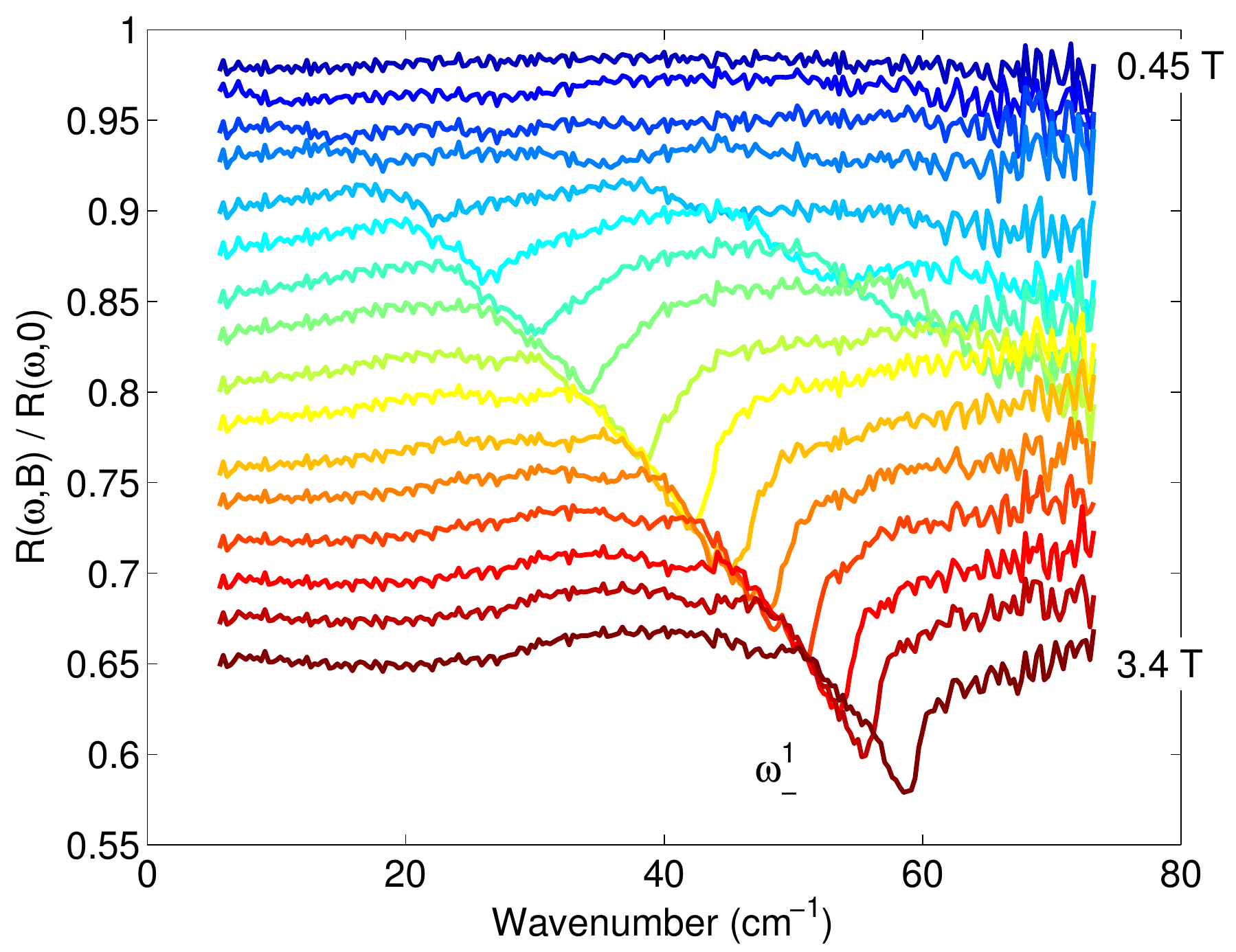}
\caption{(Color online) Time-domain measurement of the normal-incidence reflectivity of polycrystalline bismuth at 4.2~K at various magnetic fields. Ratio of reflectivity in field divided by reflectivity at zero field are plotted. Curves are shifted for clarity.}
\label{ratio_Thz}
\end{figure}

Figure~\ref{ratio_Thz} shows the ratio of the reflectivity in finite field divided by the reflectivity at zero field $R(\omega,B)/R(\omega,0)$, measured using time-domain techniques at 4.2~K up to 3.4~T. The absorption corresponding to the mode $\omega_-^1$ is very clear. The magnetic field dependence of the magneto-plasma frequencies of Figs. 2 and 3 is summarized in Fig.~\ref{wpwm} for T~=~5~K and 50~K. We see that for $B \rightarrow 0$ the frequencies of the two 'acoustic' modes are proportional to $B$.

\begin{figure}[t]
\centering
\includegraphics[width=\columnwidth]{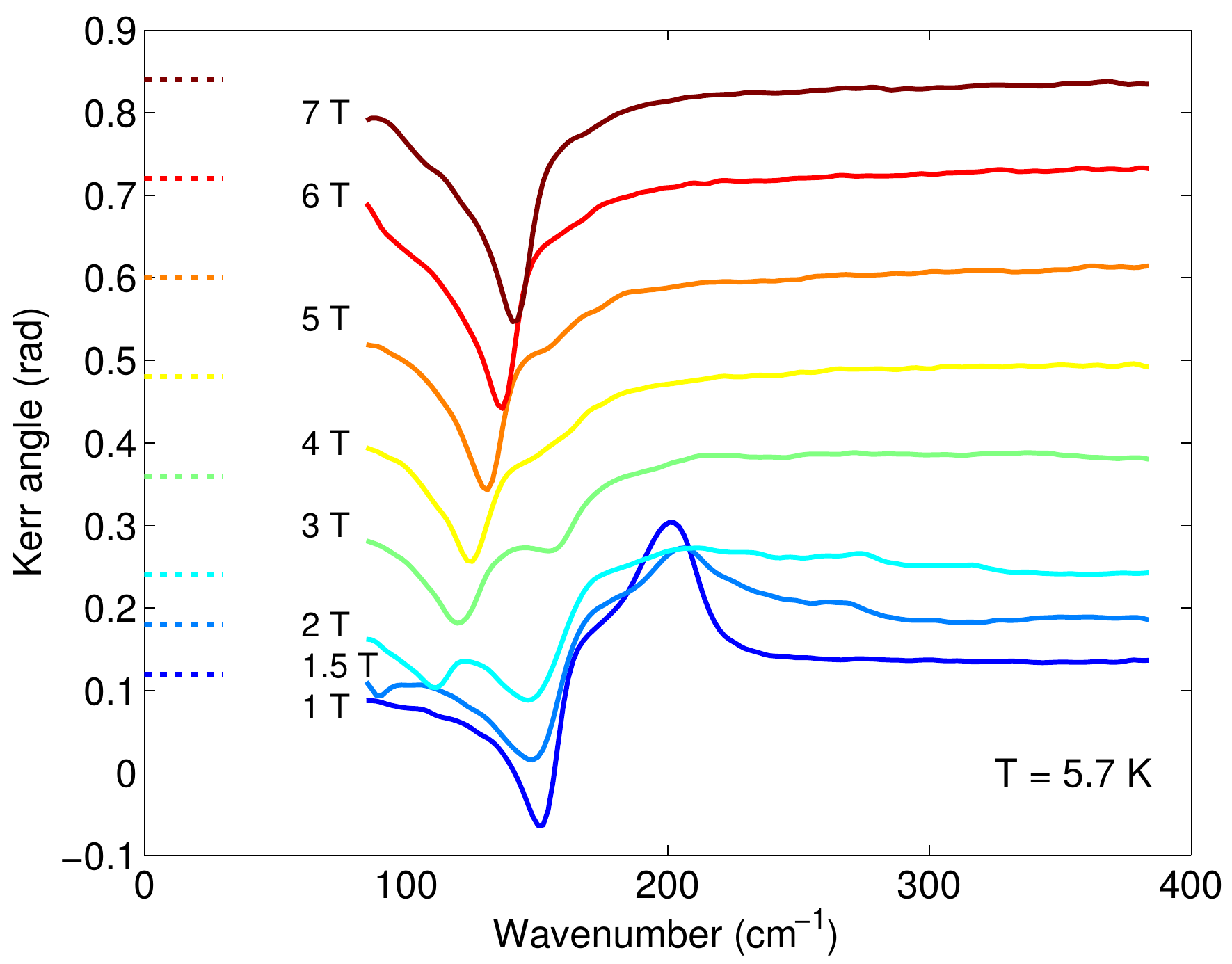}
\caption{(Color online) The Kerr angle spectra of polycrystalline
bismuth measured at 5.7 K at various magnetic fields. The curves are shifted for clarity and the offsets are the dashed lines of the same colors.} \label{Kerr}
\end{figure}

The Kerr rotation of polycrystalline bismuth at 5.7~K is shown in Fig.~\ref{Kerr}. At low field ($B <$1.5~T ), the Kerr angle exhibits a sharp structure close to the plasma frequency, going from a negative minimum to a positive maximum. The rotation is very large even at low fields, $\theta_K\simeq$ 0.2 rad at 1~T, and reaches 0.3 rad at 7~T. Moreover, it seems that the positive maximum value is linked to the plasma frequency of the second series of modes because it also disappears at 2~T. Above this field, the minimum which appears at low frequency and significantly increases in strength and to higher frequency is clearly associated to the mode $\omega_-^2$. To summarize, those observations suggest that the Kerr effect minima and maxima are linked with light and heavy carriers respectively. Therefore, an interpretation in terms of magnetoplasmon excitations fits the Kerr results at least qualitatively (see discussion).


\section{Discussion}\label{sec:theo-disc}

The use of polycrystalline sample has helped us to observe experimentally the modes $\omega_{-}^{1}$ and $\omega_{-}^{2}$ as a function of magnetic field by virtue of the fact that the grain-boundaries of the polycrystalline material make the optical oscillator strength to these modes finite. This advantage comes with a price, since the interpretation of the optical properties of polycrystalline materials is less straightforward than for single crystals. We start with the empirical observation that infrared reflectance spectra of poly-crystals usually look very much like a superposition of reflectance spectra taken with the electric field along the optical axis of the anisotropic material. In the present case the materials should be regarded as a clustering of randomly oriented crystallites. In our sample the size of a monocrystal grain is much smaller than the scattered light wavelength as well as the size of the area probed with light and the light penetration depth. So we measure the response of a 3D set of grains simultaneously. Since the grain-boundaries provide resistive barriers for electrical transport, the appropriate theoretical model is that of a resistive network of crystallites (which can be represented by $LC$-oscillators with plasma-frequency $\omega_p = \sqrt{LC}$) connected by finite resistances representing the boundaries. The resonance frequencies of such a circuit are large and given by the  plasma-frequencies of the individual grains, {\em i.e.} the highest one along the trigonal axis and the lowest frequency for the two axis perpendicular to it. For brevity we will employ here the labels $a$, $b$ and $c$ to indicate the binary, bisectrix and trigonal axis respectively.  Electromagnetic waves are resonantly scattered from a crystallite when the frequency matches one of the two plasma-resonances, with intensity proportional to the projection of the dipole moment of plasma-resonance on the electromagnetic field polarization. Another notable experimental attempt, where several plasmon branches were observed, was done on fume of Bi microparticles~\cite{Sherriff}. There is a crucial difference with respect to our sample as in a fume the inter-grain coupling is negligible.

In the data in Fig.~\ref{data}a the minima for $B=0$ occur at 192 cm$^{-1}$ and 170 cm$^{-1}$, {\em i.e.}  very close to the minima in the reflectivity of single crystalline bismuth\cite{boyle1958} for the electric field polarized along the trigonal axis (187 cm$^{-1}$) and perpendicular to the trigonal axis (158 cm$^{-1}$).  The Mie-resonances of an individual crystallite does not exactly coincide with the bulk plasma-frequencies. Factors which play a role are: (i) influence of crystallite shape on the plasma-modes, (ii) surface states and (iii) electron energy quantization due to small crystallite size. In addition, finite coupling between crystallites affects the overall optical response. It is therefore natural to interpret the reflectance minima in Fig. 2 as the plasma minima of plasmons oscillating along the trigonal axis and perpendicular to it. 

\begin{figure}[t]
\centering
\includegraphics[width=0.5\columnwidth]{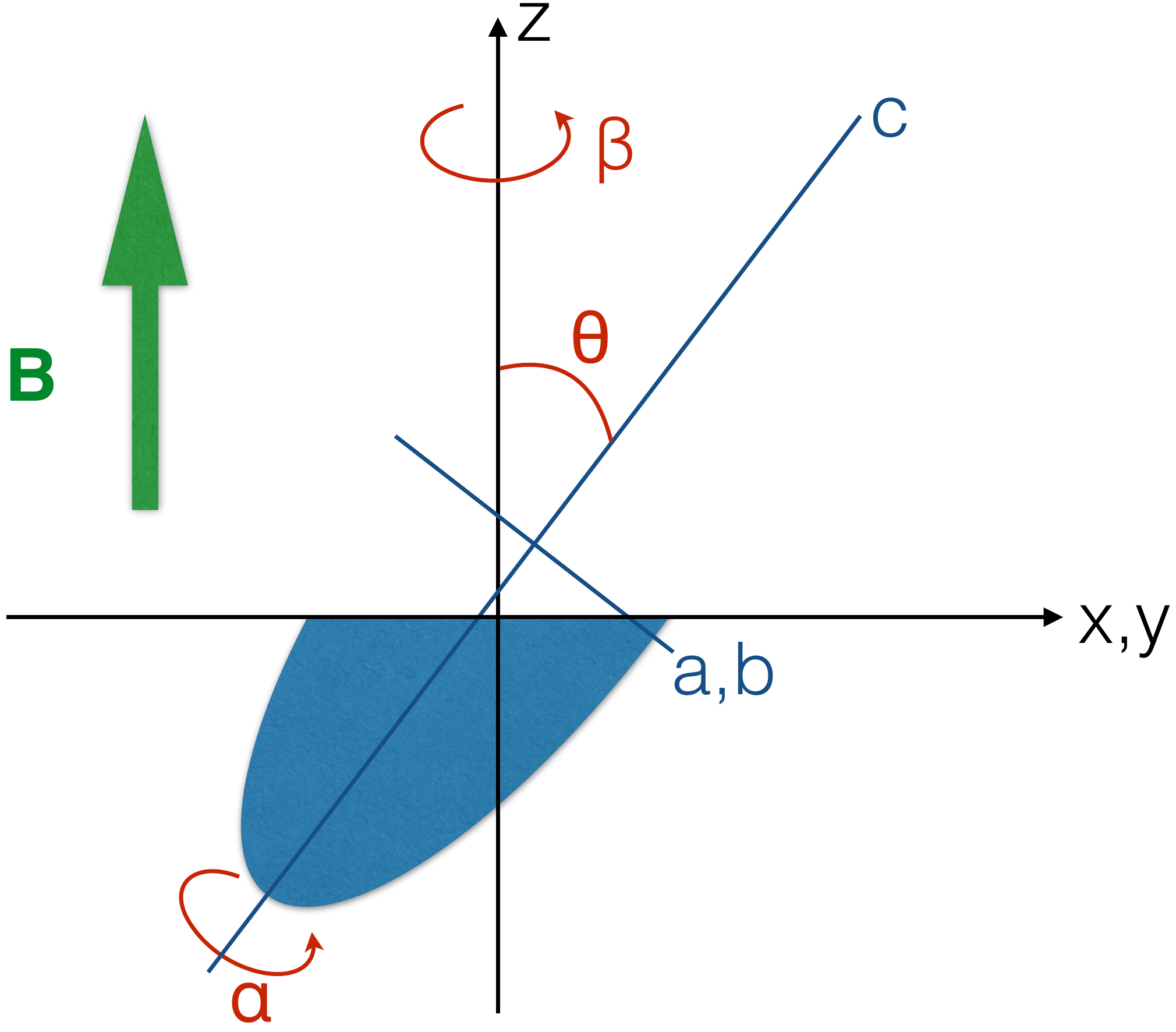}
\caption{(Color online) Geometry of a crystal grain (blue ellipse), truncated by the $x-y$ plane, in order to explain our magnetoplasmon resonances (see text).} 
\label{grain}
\end{figure}

A more quantitative microscopic description of the magnetoplasmons requires further consideration. The problem can be sketched as follows (see Fig.~\ref{grain}): the sample surface defines the $x-y$ plane, the field $B$ is applied parallel to $z$. The incoming and reflected light propagates along the $z$-axis, with the electric field in the $x-y$ plane. The $c$-axis of the individual crystallites are tilted an angle $\theta$ away from the $z$-axis. The vacuum/grain interface is in the $x-y$ plane, and is at an angle $\pi/2 -\theta$ with the $c$-axis of the crystal structure. The angles $\alpha$ and $\beta$ describe the orientation of the $c$-axis relative to the $x-y$ frame, and the rotation of $a$ and $b$ relative to the $x-y-z$ frame. The magnetic field $B$ induces a Lorentz-force on each electron, which points on the $x-y$ plane. By virtue of the anisotropy of the ellipsoidal electron and hole Fermi-surfaces this force depends on $\theta$, $\alpha$ and $\beta$.  For $B=0$ the solutions are plasma-resonances parallel to the $a-b$ plane and parallel to $c$ of the individual crystallites, with frequencies close to the bulk plasma-frequencies. 

For $B \neq 0$ the situation is rather complex, and the resonance frequencies will depend on $\theta$, $\alpha$ and $\beta$ as well as on $B$. By virtue of having both holes and electrons, acoustic and optic solutions will be found, and there will be different solutions according to the different modes of plasma-oscillation. In our optical experiment, the behavior of a large ensemble of grains is observed, averaged over all grain orientations described by $\theta$, $\alpha$ and $\beta$, and taking into account the inter-grain coupling. This latter is due to long range Coulomb interactions between carriers, superposition of Mie electric fields on grain interfaces and resistivity of inter-grain boundaries. The last two in particular can introduce a substantial imaginary part to the resonance frequencies and thus may cause overdamping of some modes (a phenomenon well known for instance in the Maxwell-Garnett approximation).

Although this constitutes a complicated problem, the fact of having electrons and holes causes the most striking phenomenon: as already mentioned in the introduction, when a system consisting of two types of carriers with different cyclotron masses (heavy and light) is subjected to a perpendicular magnetic field (Faraday geometry), two magnetoplasmon edges are expected. Their frequencies are given by~\cite{Noto74, Piotr11, Sarma83}:
\begin{align}
\omega_\pm^2  =  & \frac{1}{2}[(\omega_{p,l}^{2}+\omega_{p,h}^{2}+\omega_{c,l}^2+\omega_{c,h}^2) \nonumber \\
                                & \pm\sqrt{(-\omega_{p,l}^{2}+\omega_{p,h}^{2}-\omega_{c,l}^2+\omega_{c,h}^2)^2+4\omega_{p,l}^{2}\omega_{p,h}^{2}}]
\label{plasmon}
\end{align}
in the limit $q\rightarrow 0$, where $l$ and $h$ denotes light and heavy respectively (which correspond to electron or holes depending on the field orientation), $\omega_{p,l/h} = \sqrt{\frac{n_{l/h}e^2}{m_{l/h}\varepsilon_0\epsilon_\infty}}$ and $\omega_{c,l/h} = \frac{eB}{m_{c,l/h}}$ are the screened plasma and the cyclotron frequencies respectively and $n$ the density of carriers, $m$ the mass, $\varepsilon_0$ the vacuum permittivity and $m_c$ the cyclotron mass. $\varepsilon_\infty$ takes into account the high frequency interband transitions ($\varepsilon_\infty$ $\simeq$ 100 in Bi for E$\|ab$). This formula is obtained by resolving the equation $\varepsilon^{RPA}(q\rightarrow 0,\omega)$ = 0. Nevertheless, in our case, even with the Kerr angle, it was not possible to obtain a satisfactory dielectric tensor as a function of magnetic field\cite{Levallois13}. This is why we consider the minima of the reflectivity edges as the magnetoplasmon resonances. At zero field, $\omega_+$ is nothing else than the classical plasma frequency, equal to $\sqrt{\omega_{p,l}^{2}+\omega_{p,h}^{2}}$. At very high magnetic fields, when the cyclotron motion dominates, these two modes $\omega_{\pm}$ correspond to oscillations of heavy ($\omega_-$) and light ($\omega_+$) carriers. In the case of bismuth, the heavier (or lighter) carrier depends on sample orientation according to the magnetic field.  


\begin{figure}[t]
\centering
\includegraphics[width=\columnwidth]{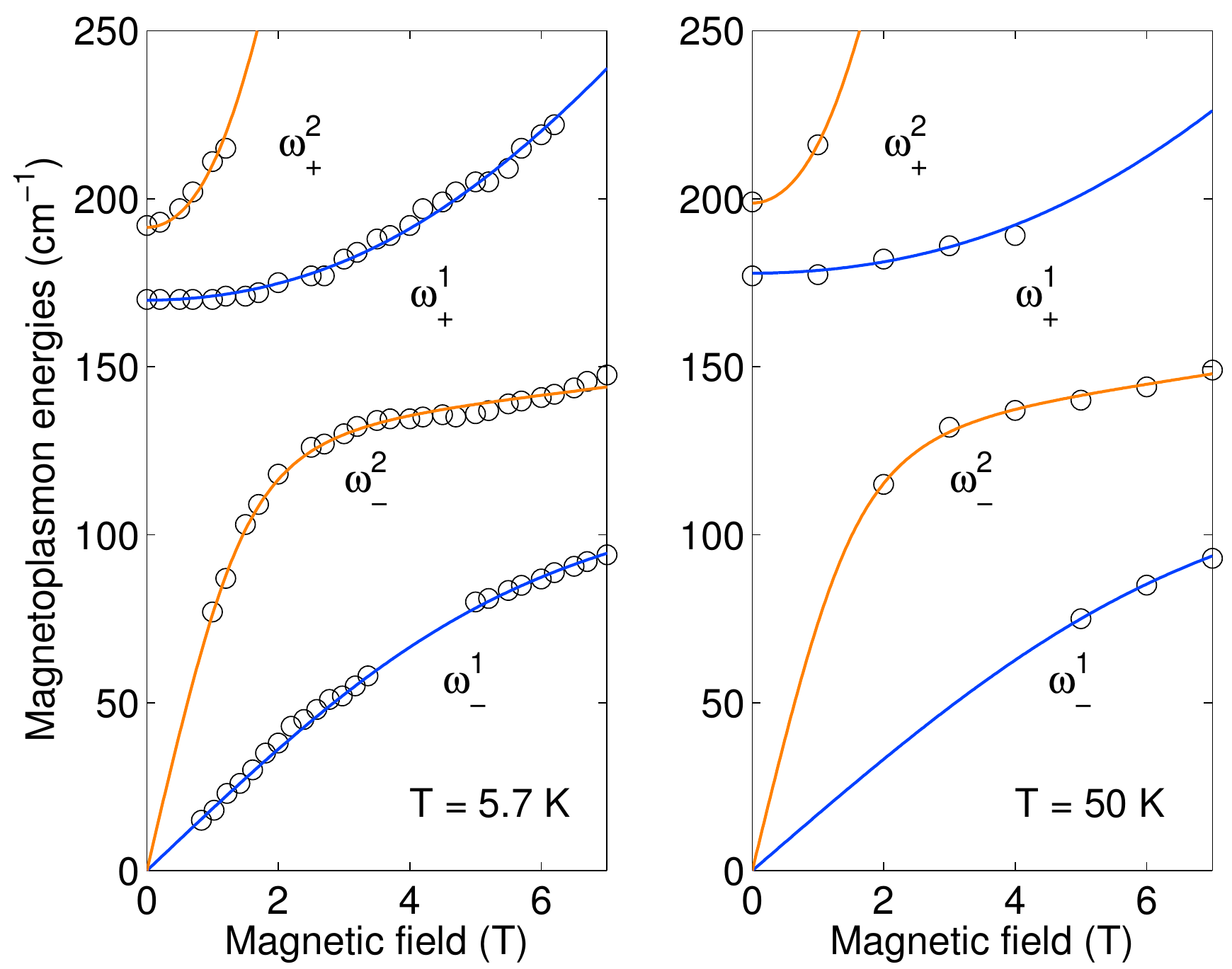}
\caption{(Color online) Magnetic field dependence of the observed collective modes $\omega_\pm$ at 5.7~K (left) and at 50~K (right). Solid lines are fit from the Eq.~\ref{plasmon}.}
\label{wpwm}
\end{figure}

We have fit Eq.~\ref{plasmon} to our experimental magnetoplasmon frequencies summarized in Fig.~\ref{wpwm}. We clearly see the two sets of modes denoted by the number 1 or 2. It is important to note that those combinations of absorptions are the only ones that can be fit with Eq.~\ref{plasmon}. The obtained parameters, the cyclotron masses and the screened plasma frequencies, are presented in Table~\ref{resultfit}. Given the good quality of the fits, we can make two observations: (i) the values of plasma frequencies are close to the ones measured for a single crystal at $B=0$~\cite{boyle1958,Tediosi,Laforge10} and (ii) the temperature dependence of $\omega_+^{1}$ is similar to determined for monocrystal~\cite{Armitage10}, which is expected as it comes from the variations of the total density when temperature changes. It appears that $\omega_-$ has no temperature dependence between 5.7 and 50~K, which is expected in low field (i.e. below the quantum limit)~\cite{Piotr11}. If we compare the very low field part of the fits of the magnetoplasmon resonances at 5.7~K with the resonance observed in microwave absorption obtained by Smith \etal~\cite{Smith63} along the three main axes of the unit cell of bismuth, we can see that those measurements, performed with the Azbel'-Kaner method ($\mathbf{B}$//surface of the sample, normal incidence of light), revealed oscillations of charge density for very similar values of magnetic fields.


\begin{table}[h!]
\caption{Fit parameters obtained with the Eq.~\ref{plasmon}
at 5.7~K and 50~K for each series of modes. Masses are in units of
the mass of the electron and optical modes are in units of
cm$^{-1}$.}
\begin{center}
\begin{tabular}{c c c c}
\hline
\hline
 Temperature &  &  series 1 & series 2\\
\hline
  5.7~K & $\omega_{p,h}$ & 115 & 139\\
         & $\omega_{p,l}$ & 125 & 132\\
         & m$_{c,h}$ & 0.275 & 0.148\\
         & m$_{c,l}$ & 0.034 & 0.008\\
  \hline
  50~K & $\omega_{p,h}$ & 125 & 141\\
         & $\omega_{p,l}$ & 127 & 140\\
         & m$_{c,h}$ & 0.424 & 0.128\\
         & m$_{c,l}$ & 0.039 & 0.008\\
\hline
\hline
\end{tabular}
\label{resultfit}
\end{center}
\end{table}


\section{Conclusion}
\label{Conclusion}

In summary, we have measured the far-infrared reflectivity of polycrystalline bismuth in magnetic field. We attribute to magnetoplasmon resonances the several features that we observed under magnetic field, which become, in the low field limit, optical and acoustic resonances of an electron-hole system. The use of polycrystalline sample has helped us to observe experimentally the acoustic plasmon modes because the translational symmetry is broken. In the limit of zero field, however, the optical modes of these two series converge onto plasma frequencies measured in monocrystal along the main optical axes. This finding, of well defined resonances, paves the way for optical spectroscopy research of materials where only polycrystalline samples are available. Valuable information, at least about the temperature or magnetic field dependence of the resonant frequencies, can be gained. The uniqueness of bismuth is probably related to extreme anisotropy of its electronic properties, thus strongly anisotropic materials would be potential candidates for further studies.

\subsection*{Acknowledgments}
The authors thank Adrien Stucky, Damien Stricker, Micha\"el Tran and Radovan Cerny for technical assistance and are grateful to Thierry Giamarchi for fruitful
discussions. This work is supported by MaNEP and by the SNSF through Grant No. 200020-135085 and the National Center of Competence in Research
(NCCR) "Materials with Novel Electronic Properties-MaNEP".

\appendix

\section*{Appendix}

\section{Standard theory of
magnetoplasmon}\label{app:theory-standard}

The collective mode (we are interested in) is defined as a zero of a longitudinal dielectric function $\epsilon_{||}(q,\omega)$ of a material. An effective interaction in any material is a bare Coulomb interaction $V_0(q)$ divided by a tensor $\hat{\epsilon}(q,\omega)$, the $\epsilon_{||}(q,\omega)$ being
its diagonal component. The dielectric function contains all screening processes and is build out of a series of particle-hole bubbles. The result of the simplest random phase (RPA) re-summation of such scattering processes gives an effective interaction:
\begin{equation}\label{eq:RPA-inter}
    V_{eff}(q,\omega)=\frac{V_{0}(q)}{\epsilon_{||}(q,\omega)}=\frac{V_{0}(q)}{1-V_{0}(q)\sum_{\nu}\Pi_{\nu}(q,\omega)}
\end{equation}
where the sum is taken over polarizabilities $\Pi_{\nu}(q,\omega)$ of all families $\nu=h,e1,e2,e3$ of carriers in bismuth. The denominator of Eq.~\ref{eq:RPA-inter} defines $\epsilon_{||}(q,\omega)$. The polarizability $\Pi_{\nu}(q,\omega)$ is defined as a Fourier transform of time ordered correlation function of densities: $\Pi_{\nu}^{xx}(q,\tau)\equiv -\langle T_{\tau} \rho_{\nu}^{x}(q,\tau)\rho_{\nu}^{x}(-q,0)\rangle$ (where for once we put the space index, to emphasize that we work with longitudinal polarization, with the electric field $\vec{E} \parallel x$). In the case of non-interacting fermions without magnetic field, the polarizability is given by the so called electron-hole (two particle) propagator in the pocket $\nu$:

\begin{equation}\label{eq:polarizability-Lindhard}
     \Pi_{\nu}^{0}(q,\omega)=\sum_{\vec{k}}\frac{f(\xi_k^{\nu})-f(\xi_{k+q}^{\nu})}{\omega+\xi_{k}^{\nu}-\xi_{k+q}^{\nu}+0^{+}}
\end{equation}

\noindent where we did not include the spin index, a summation over it is
implicit (bismuth is a non-magnetic material), and $f(\xi_k^{\nu})$ is
a Fermi-Dirac distribution. The masses enter to our problem
through \emph{eigen-energies} $\xi_{\vec{k}}^{\nu} \sim
\sum_{a,b,c} k_{i}^2/2m^{\nu}_{i}$ of carriers in each pocket
$\nu$~\footnote{$m_{a,b,c}$ are masses along the axis of
ellipsoidal pocket $\nu$. In Bi these masses are extremely
anisotropic, and in each ellipsoid main axes are oriented
differently, see Fig.\ref{Bismuth}.}. The sum in
Eq.~\ref{eq:polarizability-Lindhard} is known and leads to the
Lindhard function. Eq.~\ref{eq:polarizability-Lindhard} is
useful for us throughout the discussion, the principal
Eq.~\ref{plasmon} can be derived using
$\Pi_{\nu}^{0}(q_{x}\rightarrow 0,\omega)$ (for simplistic model
of two spherical pockets and if the magnetic field energy is
included in the \emph{eigen-energies}).

Magnetic field $\vec{B}$ (with certain orientation $\parallel z$,
taken perpendicular to $\vec{E}$) enters via cyclotron mass to the
eigen-energies $\xi_k^{\nu}$ and thus affects the polarizability,
which is now $\Pi_{\nu}(q,\omega; \vec{B})$. The cyclotron mass
depends on the cross section (in the plane $x-y$) of a given
pocket of a Fermi surface, thus in the case of bismuth it is
different for different carriers.

In a polycrystal, when size of grains is small, the formula Eq.~\ref{eq:polarizability-Lindhard} is 
slightly modified: due to a finite size the allowed values of momenta are quantized and if carriers can propagate between grains then 
we must allow for inter-grain scattering, an effect which enters inside a bubble as a vertex correction. Moreover some screening processes
may take place in the neighboring grains, which inserts additional bubbles, with different masses, into the RPA series. This is how an inter-grain coupling enters into plasmon dispersion.

The density-density interactions which drive the (optical) plasmon instability are of a long range order. This is because screening is caused by plasmons themselves so interactions are screened only for processes with energy lower than the frequency of collective excitations. The fact that the electron-electron interactions have a long range character means that the density-density  interactions dominate. Moreover large anisotropy of a mass tensor implies that only a limited fraction of interfaces are transparent for the electron-hole waves (in fact, due to vertex corrections, also within $\Pi_0$ itself the low angle scattering is preferable). In these circumstances, when we resort to pole expansion of $\epsilon_{||}$ in Eq.~A1, the quasi-classical approximation of coupled harmonic oscillators, used in the Discussion, is applicable.

\section*{References}
\bibliographystyle{apsrev4-1}
\bibliography{biblio}

\end{document}